\documentclass[
letter,
twocolumn,
superscriptaddress,
amsmath,
amssymb,
prl,
nofootinbib
showkeys,
10pt,
floatfix,
nobibnotes,
aps,
fltpage
]{revtex4-2}

\usepackage{latexsym}
\usepackage{amsmath}
\usepackage{amssymb}
\usepackage[T1]{fontenc}
\usepackage[open]{bookmark}
\usepackage{hyperref}
\hypersetup{colorlinks=true,allcolors=blue}
\usepackage{upgreek}

\usepackage{orcidlink}

\usepackage{dcolumn}
\usepackage{bm}
\usepackage{xcolor}
\usepackage{hyperref}
\usepackage{cleveref}
\usepackage{float}
\usepackage{svg}
\usepackage{amssymb}
\usepackage{scalerel}
\usepackage{caption}
\usepackage{wasysym}
\usepackage{lipsum}
\usepackage{enumitem}

\providecommand{\pnl}[1]{{\textcolor{black}{(#1)}}}

\makeatletter
\pretocmd\frontmatter@keys@format{\addvspace{20\p@}}{}{}
\makeatother

\usepackage{svg}
\usepackage{cleveref}
\usepackage{siunitx}
\usepackage{caption}
\usepackage{multirow}
\usepackage{makecell}
\usepackage{tabularray}
\usepackage{makecell} 
\usepackage{array} 
\usepackage{graphicx} 
\usepackage{booktabs} 
\newcommand{\polima}{POLIMA---Center for Polariton-driven Light--Matter Interactions, University~of~Southern~Denmark, Campusvej 55, DK-5230 Odense M, Denmark}
\newcommand{\dias}{D-IAS---Danish Institute for Advanced Study, University~of~Southern~Denmark, Campusvej 55, DK-5230 Odense M, Denmark} 
\newcommand{\kobe}{Department of Electrical and Electronic Engineering, Kobe~University, Rokkodai, Nada, Kobe 657-8501, Japan} 

\begin{document}

\title{Probe- and Substrate-Dependent Visibility of Mie Resonances in Silicon Nanospheres}

\author{Yonas Lebsir\,\orcidlink{0000-0002-9383-0278}}
\affiliation{\polima}
\author{Huatian Hu\,\orcidlink{0000-0001-8284-9494}}
\affiliation{\polima}
\author{P.~A.~D.~Gon\c{c}alves\,\orcidlink{0000-0001-8518-3886}}
\affiliation{\polima}
\author{Hiroshi Sugimoto\,\orcidlink{0000-0002-1520-0940}}
\affiliation{\kobe}
\author{Minoru~Fujii\,\orcidlink{0000-0003-4869-7399}}
\affiliation{\kobe}
\author{N. Asger Mortensen\,\orcidlink{0000-0001-7936-6264}}
\affiliation{\polima}
\affiliation{\dias}
\author{Christos~Tserkezis\,\orcidlink{0000-0002-2075-9036}}
\affiliation{\polima}
\affiliation{\dias}
\author{Sergii Morozov\,\orcidlink{0000-0002-5415-326X}}
\email[Corresponding author:~]{semo@mci.sdu.dk}
\affiliation{\polima}

\date{\today}

\begin{abstract}
Silicon nanospheres are high-quality optical resonators and promising building blocks for Mie-tronic devices. While the Mie resonances of an isolated sphere are well understood, practical implementations require substrates that inevitably modify the measured optical response. Here, we investigate how substrates alter the observable spectrum of individual nanospheres, focusing on three fundamentally different cases: a thin silicon nitride membrane, that emulates a free-standing particle, bulk silicon, which is common in experiments, and gold, where mirror charges lead to hybrid optical modes. Cathodoluminescence and dark-field spectroscopy, combined with electrodynamic simulations, show that the measured resonances are not intrinsic to the particle but depend strongly on the environment and the excitation mechanism. We find that substrate-induced effects and probe-specific selection rules can suppress, enhance, or even invert the spectral signatures of electric and magnetic modes. These results provide practical guidelines for interpreting and designing substrate-supported dielectric resonators for Mie-tronic applications.
\end{abstract}

\maketitle

\section{Introduction}
Subwavelength particles of high-refractive-index dielectrics support Mie resonances~\cite{Mie1908} whose electric and magnetic character can be controlled through particle size and material choice~\cite{etxarri_oex19,Kuznetsov2016, Jahani2016,verre_natnano14}. Silicon (Si) nanospheres exhibit coexisting electric dipole (ED) and magnetic dipole (MD) modes~\cite{Evlyukhin2012, Kuznetsov2012,tang_natcom14}, originating from the excitation of polarization charges with different phases across the particle, whose interference enables directional scattering~\cite{Kerker1983, Fu2013, Person2013} and underpins Huygens’ metasurfaces, meta-optics, and low-loss light–matter interaction~\cite{Decker2015, Kruk2017, Caldarola2015}. These concepts form the basis of Mie-tronics, where dielectric nanoresonators act as building blocks for subwavelength photonic devices~\cite{Kivshar2022,rybin_npjnanoph1}, including applications in lasing~\cite{mylnikov_nn14}, nonlinear optics~\cite{koshelev_sci367}, or magnetic resonance imaging~\cite{shchelokova_natcom11}.

Silicon nanospheres are a convenient platform for Mie-tronics due to their high material quality and reproducible geometry. Advances in synthesis enable colloidal crystalline Si spheres with near-perfect shape and well-controlled size~\cite{Sugimoto2020}, making them suitable model systems for studying Mie resonances in realistic device configurations.
For isolated spheres in free space, Mie theory provides exact predictions~\cite{Mie1908,Bohren1998}. In realistic implementations, however, nanoparticles (NPs) are placed on substrates that modify the modal landscape by shifting resonances, lifting degeneracies, and reshaping radiation patterns~\cite{vandeGroep2013, Miroshnichenko2015, Staude2017}, thus affecting the coupling with nearby dipole emitters, for which Si spheres are promising templates~\cite{stamatopoulou_osac4,siegel_smalls6}. While these effects are well documented for far-field observables, it is less clear to what extent the measured optical spectrum depends on the excitation mechanism.

Dark-field (DF) microscopy probes the far-field scattering cross section and is widely used to identify Mie modes in single NPs~\cite{Evlyukhin2012, Sugimoto2017, Sugimoto2020, Hinamoto2021}. Cathodoluminescence (CL) microscopy provides a complementary nanoscale probe in which a focused electron beam excites the particle and the emitted light depicts the radiative local density of optical states (LDOS)~\cite{GarciadeAbajo2010, Losquin2015, Polman2019}. In Si NPs, CL reveals mode-selective excitation governed by spatial overlap with modal fields~\cite{Coenen2013}, resolves higher-order multipoles~\cite{Matsukata2019}, and shows a dependence on electron-beam energy~\cite{Soler:2025}. Other electron-beam techniques such as electron energy-loss spectroscopy (EELS) probe both radiative and non-radiative modes~\cite{GarciadeAbajo2010, Polman2019}, while near-field optical methods access surface fields but not internal displacement currents~\cite{Matsukata2019}. A central challenge is that different experimental probes access different physical observables and excitation pathways, which can lead to conflicting or even misleading interpretations of the same mode. In electron-beam spectroscopies, in particular, mechanisms related to the electrons themselves, such as transition radiation and Cherenkov radiation further complicate the picture~\cite{stamatopoulou_josab42}. As a result, the measured spectrum does not directly reflect the intrinsic resonances of the NP but is entangled with how the modes are excited and detected. This ambiguity becomes a critical limitation for Mie-tronics, where device performance relies on precise control of modal content in realistic, substrate-supported geometries.

Here, we perform DF and CL spectroscopy on the same individual colloidal crystalline Si nanospheres~\cite{Sugimoto2020} on three distinct substrates: silicon nitride (Si$_3$N$_4$), silicon (Si), and gold (Au). These substrates, each characterized by different optical response and coupling capability, provide controlled conditions to examine how the presence of a supporting medium perturbs the modal structure of dielectric nanoresonators. We combine experiment with Mie theory and numerical simulations to analyze the observed optical response. This approach allows us to assess the advantages and limitations of DF and CL for probing Mie resonances, and to clarify how substrate and excitation influence the measured spectra. In particular, we show how and why mode profiles observed in CL can deviate from those expected from idealized theoretical descriptions.

\section{Results and discussion}

\subsection{Isolated-sphere reference}

\begin{figure}[ht]
\includegraphics[width=0.99\linewidth]{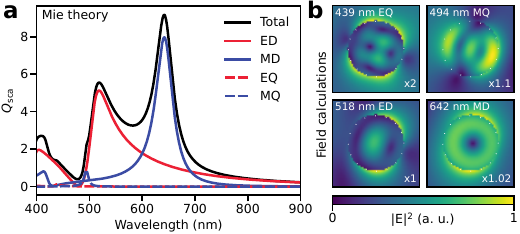}
\caption{\textbf{Mie modes in an isolated Si sphere.} 
\textbf{\pnl{a}} Mie-theory scattering efficiency ($Q_\mathrm{sca}$) of a $D=160$\,nm Si sphere in vacuum, decomposed into ED, MD, EQ, and MQ contributions. 
\textbf{\pnl{b}} Near-field intensity distributions ($|E|^2$) for each mode under unpolarized excitation; multiplication factors indicate relative scaling.}
\label{fig:fig1}
\end{figure}

To place the substrate-dependent measurements in context, we first establish an idealized substrate-free reference for a diameter $D=160$\,nm crystalline Si sphere. Using analytical Mie theory~\cite{Bohren1998} together with measured optical constants~\cite{Green1995}, we compute the scattering efficiency and modal near-field distributions of an isolated sphere. The multipolar decomposition in Fig.~\ref{fig:fig1}\pnl{a} identifies four modes in the visible spectral range: the magnetic dipole (MD, $\lambda\sim 642$\,nm) is the dominant resonance, followed by the electric dipole (ED, $\lambda\sim 518$\,nm), with weaker electric and magnetic quadrupole contributions (EQ, $\lambda\sim 439$\,nm; MQ, $\lambda\sim 494$\,nm)~\cite{Evlyukhin2010, Liu2018}. Importantly for the experiments, at this particle size the Mie resonances lie in a spectral region where all three substrates exhibit a relatively featureless CL response~\cite{Ebel:2025}.

The corresponding near-field intensity distributions in Fig.~\ref{fig:fig1}\pnl{b} reveal a distinction that is central for interpreting the substrate effects. The MD electric field is concentrated predominantly inside the NP, whereas the ED electric field extends substantially beyond the NP boundary. Because coupling to a substrate is mediated by the external electric field, modes with stronger field intensity at the NP--substrate interface are expected to experience stronger environmental perturbation. This immediately suggests that the ED should be more sensitive to substrate-induced spectral shifts and dissipation channels, while the MD, owing to its weaker external electric field, should remain comparatively robust. 

\subsection{Mode probing}

\begin{figure}[ht]
\includegraphics[width=0.99\linewidth]{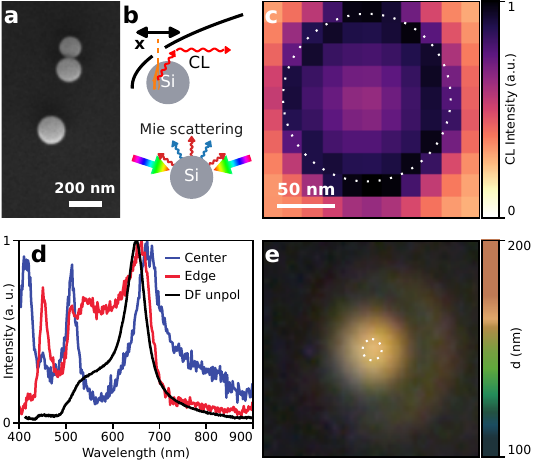}
\caption{\textbf{Mode probing in Si nanospheres.} 
\textbf{\pnl{a}}~SE image of colloidal crystalline Si nanospheres. 
\textbf{\pnl{b}}~Schematics of CL and DF excitation geometries.
\textbf{\pnl{c}}~Spectrally integrated CL map of a 160\,nm sphere. The dashed white circle indicates the edge of the sphere.
\textbf{\pnl{d}}~Far-field DF spectrum (black line) and CL spectra acquired in the center (blue line) and edge (red line) of the same 160\,nm sphere.
\textbf{\pnl{e}}~DF image of the same sphere; colorbar calibrated from 10 particles spanning 100--200\,nm diameter. The dashed white circle indicates the edge of the sphere.
}
\label{fig:fig2}
\end{figure}

The experimental platform and mode-probing techniques are summarized in Fig.~\ref{fig:fig2}. Fig.~\ref{fig:fig2}\pnl{a} shows a secondary-electron (SE) image of colloidal Si nanospheres with diameters around 160\,nm. The particles are crystalline, nearly perfectly spherical, and were synthesized following the procedure reported in Ref.~\cite{Sugimoto2020} (see Methods).

We introduce the capabilities of DF and CL probing techniques using the example of an isolated $D=160$\,nm Si sphere placed on a thin Si$_3$N$_4$ membrane of thickness $w=5$\,nm. 
As we show in the next subsection, this configuration resembles very closely the case of a free-standing NP, and thus serves as the ideal starting point for any experimental study, where some supporting medium is unavoidable.
Fig.~\ref{fig:fig2}\pnl{b} illustrates the two complementary excitation schemes used to probe the optical response. 
In CL measurements, a focused electron beam excites the NP, and the emitted light—collected by a high–numerical-aperture parabolic mirror—reflects the LDOS with nanometer-scale spatial resolution~\cite{GarciadeAbajo2010, Coenen2017}. This sub-diffraction-limited excitation makes it possible to probe how the response depends on beam position (impact parameter $b$) within a single sphere. The panchromatic CL map acquired for an electron beam accelerated at 30\,keV in Fig.~\ref{fig:fig2}\pnl{c} illustrates this capability: each pixel contains full spectral information, while the displayed intensity corresponds to the wavelength-integrated signal. Examples of the CL spectra from the edge and center of the sphere are shown in Fig.~\ref{fig:fig2}\pnl{d} by red and blue lines, respectively. 

CL couples to modes through the $z$-component of the electric field along the beam trajectory---not through the magnetic field, even for the MD~\cite{GarciadeAbajo1999, Stamatopoulou2024}. The electric field of the MD [Fig.~\ref{fig:fig1}\pnl{b}], arising from the circulating displacement currents that are caused by retardation-affected excitation of polarization charges, peaks near the particle surface rather than at the center where the magnetic field is maximal. Both ED and MD are therefore excited most efficiently at peripheral beam positions~\cite{Soler:2025, Matsukata2019}, as confirmed by the annular intensity profile in the panchromatic CL map [Fig.~\ref{fig:fig2}\pnl{c}].
Mode selectivity in CL arises from the distinct radial $E_z$ profiles: the ED ring is sharp and surface-concentrated, extending beyond the particle boundary, while the MD ring is broader and extends further inward~\cite{Coenen2013, Soler:2025}. This example highlights the ability of CL to resolve the spatial distribution of individual optical modes within a single NP. 

Fig.~\ref{fig:fig2}\pnl{b} also schematically shows the DF method, where broadband illumination probes the far-field scattering response of the NP. In contrast to CL, DF collects the response of the entire particle and does not provide spatially localized information from within a single sphere. It can, however, be extended to polarization-resolved excitation and detection~\cite{Sinev2016} to selectively probe different dipolar responses, including out-of-plane electric and magnetic dipole contributions, by using the analyzer to suppress the dominant co-polarized signal and enhance sensitivity to specific mode symmetries (see Methods). The DF image of the same particle shown in Fig.~\ref{fig:fig2}\pnl{e} provides the complementary far-field signature, while the unpolarized DF spectrum is shown in Fig.~\ref{fig:fig2}\pnl{d}. 
The dark-field image in Fig.~\ref{fig:fig2}\pnl{e} shows the characteristic size-dependent Mie scattering color~\cite{Evlyukhin2012, Kuznetsov2012}. For the isolated sphere, all orientational components of each dipole mode are degenerate---confirmed by the equivalence of polarization-resolved field distributions in Fig.~\ref{fig:fig1}\pnl{b}.

The CL and DF spectra in Fig.~\ref{fig:fig2}\pnl{d} highlight the capabilities of the two probing techniques. While CL provides spatial information on mode excitation, it is also capable of exciting blueshifted higher-order modes. The two techniques therefore probe different physical observables and rely on different excitation mechanisms. DF serves as a far-field reference, whereas CL provides localized excitation that is sensitive to the electron trajectory. By combining DF and CL on the same nanosphere and across different substrates, we can separate how the environment perturbs the modes from the way each probe excites them. Together, these measurements provide a more complete picture of the modal structure than either technique alone; complementary techniques such as EELS can further probe both radiative and non-radiative modes~\cite{GarciadeAbajo2010,Polman2019}.

\begin{figure*}[t!]
\includegraphics[width=0.99\linewidth]{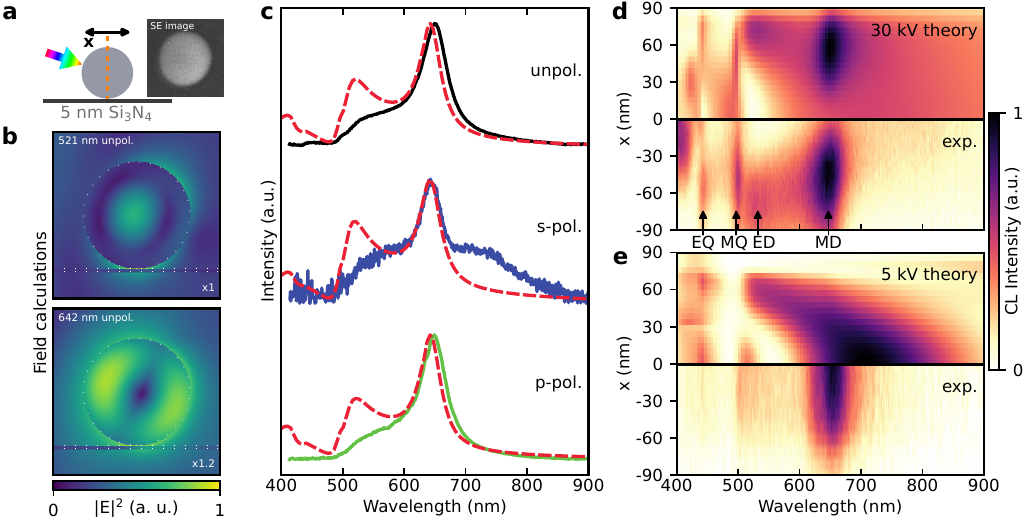}
\caption{\textbf{Particle (Si) on weakly perturbing membrane (Si$_3$N$_4$).} 
\textbf{\pnl{a}} Measurement geometry. 
\textbf{\pnl{b}} Selected near-field distributions ($|E|^2$) for ED and MD modes on the membrane. 
\textbf{\pnl{c}} Polarization-resolved DF spectra at 72$^\circ$ incidence, contrasting unpolarized (black line), s-polarized (blue line), and p-polarized (green line) with theoretical counterparts (red dashed lines). 
\textbf{\pnl{d,e}} Position-resolved CL spectra, contrasting theory (positive $x$-values) with experiments (negative $x$-values) for both 30\,kV [\pnl{d}] and 5\,kV [\pnl{e}]. At 5\,kV the ED is suppressed to a weak shoulder, whereas at 30\,kV both ED and MD are resolved, demonstrating beam-broadening-controlled mode selectivity.}
\label{fig:fig3}
\end{figure*}

\subsection{Weakly perturbing substrate: {\uppercase{S}\lowercase{i}$_3$\uppercase{N}$_4$} membrane }

We now turn to the first supported-particle geometry, namely a Si sphere placed on a $w=5$\,nm thick Si$_3$N$_4$ membrane [Fig.~\ref{fig:fig3}\pnl{a}]. This configuration is designed to approximate the isolated-sphere limit while providing an experimentally accessible platform for both DF and CL measurements. Owing to the membrane's small thickness and its somewhat moderate dielectric function of $\epsilon\sim 4$ (relative to $\epsilon=1$ for air and $\epsilon\sim 16$ for Si), it is expected to perturb the Mie modes only weakly. Furthermore, for the electron-beam voltages considered, the electron mean-free path $\ell$ satisfies $\ell \gg w$~\cite{Ebel:2025}. 

To assess the effect of the membrane in the far field, we first examine the DF spectra in Fig.~\ref{fig:fig3}\pnl{c}. They show that the supported particle remains close to the isolated-sphere limit: the unpolarized response [Fig.~\ref{fig:fig3}\pnl{c}, top] is dominated by the MD peak near $\lambda\sim 640$\,nm, with a weaker ED feature near $\lambda\sim 520$\,nm, in qualitative agreement with the Mie prediction [Fig.~\ref{fig:fig1}\pnl{a}] and the simulations. The s- and p-polarized spectra are nearly identical [Fig.~\ref{fig:fig3}\pnl{c}], indicating that the 5\,nm Si$_3$N$_4$ membrane introduces little polarization anisotropy despite the formal symmetry breaking imposed by the supporting membrane. 
However, the experimental spectra are only qualitatively captured by the simulated ones: the relative ED/MD amplitudes differ and small spectral shifts are present. Such deviations are expected for a real particle and can arise from the combined effect of the membrane, finite collection geometry, and small deviations of the experimental sphere from the idealized model.

The computed field distributions in Fig.~\ref{fig:fig3}\pnl{b} likewise remain very similar to those of the isolated sphere, and their polarization-resolved variants are in practice nearly indistinguishable, indicating that the electromagnetic overlap with the 5\,nm membrane is too weak to substantially lift the orientational degeneracy. Still, a first substrate effect is already visible: the narrow gap between the Si sphere and the membrane weakly concentrates the electric field at the particle--substrate interface.

The CL measurements at 30\,kV in Fig.~\ref{fig:fig3}\pnl{e} show that experiment and theory are in very good overall agreement. The position-resolved CL map displays the expected mode-selective excitation of the Si sphere: the spectral response varies with beam position, or impact parameter, reflecting the spatial overlap between the electron trajectory and the modal near fields~\cite{GarciadeAbajo2010,Losquin2015,Soler2025}. In particular, the edge and center trajectories excite different mixtures of ED- and MD-like response, consistent with the standard CL picture for resonant Si nanospheres on thin membranes. The ED is strongest near the particle edge ($|x|\approx 60 - 80$\,nm), whereas the MD extends further inward, in agreement with position-resolved CL studies of Si nanospheres on Si$_3$N$_4$~\cite{Fiedler:2022,Soler:2025}. 
The main discrepancy is a feature around $\lambda\sim 400$\,nm in the sphere center, that is absent in the simulations. This feature appears only when the beam passes close to the sphere center and is not reproduced by the electrodynamic model. We therefore tentatively attribute it to an additional electron-induced emission channel, most likely transition radiation, rather than to a resonant particle eigenmode~\cite{GarciadeAbajo2010,Fiedler:2022}.

\begin{figure*}[ht]
\includegraphics[width=0.99\linewidth]{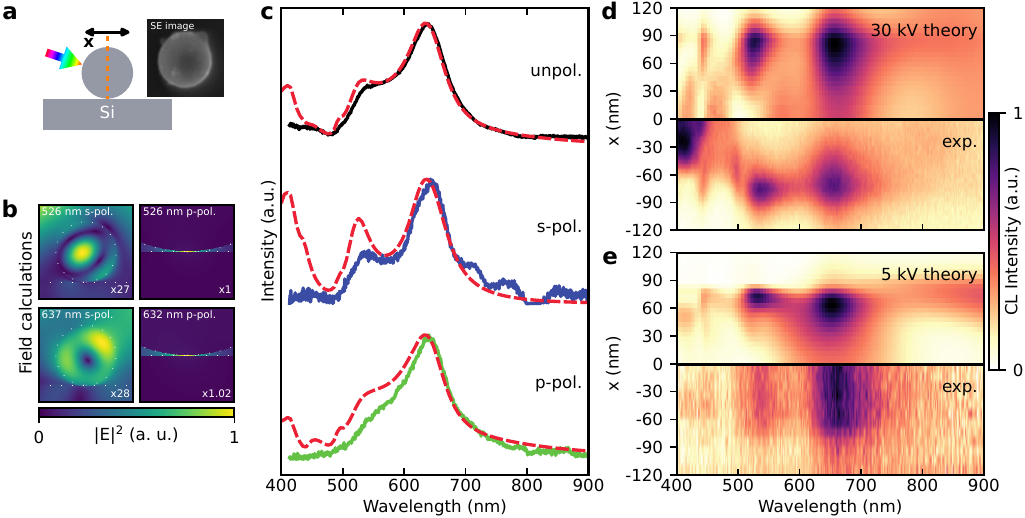}
\caption{\textbf{Particle (Si) on reflecting dielectric substrate (Si).} 
\textbf{\pnl{a}} Measurement geometry. \textbf{\pnl{b}} Selected near-field distributions ($|E|^2$) showing ED field concentration at the particle--substrate interface under s- and p-polarization. 
\textbf{\pnl{c}} Polarization-resolved DF spectra at 72$^\circ$ incidence, contrasting unpolarized (black line), s-polarized (blue line), and p-polarized (green line) excitations with theoretical counterparts (red dashed lines). Polarization-resolved DF spectra show enhanced ED/MD ratio relative to Si$_3$N$_4$ (see Fig.~\ref{fig:fig3}). 
\textbf{\pnl{d,e}} Position-resolved CL spectra, contrasting theory (positive $x$-values) with experiments (negative $x$-values) for both 30\,kV [\pnl{d}] and 5\,kV~[\pnl{e}].
}
\label{fig:fig4}
\end{figure*}

A qualitatively different picture emerges at 5\,kV [Fig.~\ref{fig:fig3}\pnl{d}]. In the experiment, the spatial selectivity seen at 30\,kV is largely lost: nearly the same spectral response is observed across the NP, indicating that the impact-parameter-dependent excitation of individual modes is no longer resolved. We attribute this to strong multiple scattering of low-energy electrons inside the Si sphere, which broadens the excitation volume and effectively averages the coupling over the particle. This is consistent with Monte Carlo simulations, which show substantially stronger beam spreading at 5\,kV than at 30\,kV. The corresponding electrodynamic simulation, however, clearly fails to reproduce the measured response. Because it still treats the electron beam as a straight line current, it does not capture the spatially broadened excitation at low energy and therefore does not reproduce the experimental lack of spatial selectivity. In the calculation, only weak residual CL remains besides the ED and quadrupolar contributions, whereas the experiment shows an additional broad signal, strongest near the particle center where the Si thickness is largest. We therefore assign this background-like emission to electron-induced Si emission, most likely involving interband transitions~\cite{Ebel:2025}. Notably, the $\lambda\sim 400$\,nm feature seen at 30\,kV is absent at 5\,kV, supporting its assignment to transition radiation or Si interband-related emission (for the latter, see Ref.~\cite{Dong2019}).
Overall, these results complement the phase-matching picture proposed for energy-dependent CL excitation of Si nanospheres~\cite{Soler2025}: at low electron energy, loss of beam collimation through inelastic scattering becomes a decisive factor that limits mode-selective excitation.

\subsection{Reflecting dielectric substrate: \uppercase{S}\lowercase{i}}

We next consider the Si sphere on bulk Si, shown in Fig.~\ref{fig:fig4}\pnl{a}. In contrast to the weakly perturbing Si$_3$N$_4$ membrane, the high-index Si substrate already produces a clear mode-selective modification of the optical response. This is evident in the unpolarized DF spectra [Fig.~\ref{fig:fig4}\pnl{c}], where the ED/MD peak intensity ratio increases substantially compared with the membrane case, indicating preferential enhancement of the ED scattering efficiency. Such behavior is physically expected, since substrate coupling increases with substrate permittivity and is strongest for modes whose fields extend into the substrate region~\cite{vandeGroep2013,Staude2017}. 
The polarization-resolved DF spectra show that the ED is more pronounced for s-polarized excitation and suppressed for p-polarization. This is consistent with polarization-resolved DF studies of Si NPs under oblique illumination, where the ED was found to be strongly polarization-sensitive, while the MD was much less affected~\cite{Permyakov2015}. More generally, for substrate-supported Si NPs the substrate modifies the coupling and radiation of different dipolar components under s- and p-polarized excitation~\cite{Kuo2016,Sinev2016}.

\begin{figure*}[t!]
\includegraphics[width=0.99\linewidth]{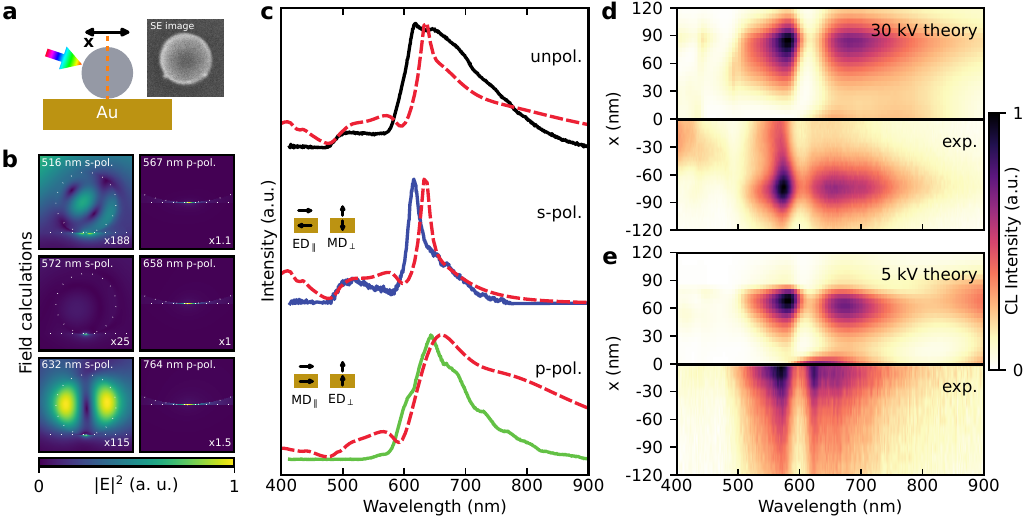}
\caption{\textbf{Particle (Si) on reflecting metallic mirror (Au).} 
\textbf{\pnl{a}} Measurement geometry. 
\textbf{\pnl{b}} Near-field distributions showing the ED gap mode under p-polarization and MD fields under s-polarization, illustrating image-dipole-induced symmetry breaking. 
\textbf{\pnl{c}} Polarization-resolved DF spectra at 72$^\circ$ incidence, contrasting unpolarized excitation (black line), s-polarized excitation of ED$_\parallel$ and MD$_\perp$ (blue line), and p-polarized excitation of MD$_\parallel$ and ED$_\perp$ (green line) with theoretical counterparts (red dashed lines). The ED appears as a featureless continuum in DF but as an enhanced resonance in CL, demonstrating that the two techniques probe distinct observables. 
\textbf{\pnl{d,e}} Position-resolved CL spectra, contrasting theory (positive $x$-values) with experiments (negative $x$-values) for both 30\,kV [\pnl{d}] and 5\,kV~[\pnl{e}].}
\label{fig:fig5}
\end{figure*}

The calculated field distributions in Fig.~\ref{fig:fig4}\pnl{b} support this picture. Under p-polarized excitation, the ED field is pulled toward the particle--substrate interface, where it becomes significantly concentrated in the gap and penetrates into the Si half-space, while the MD electric field remains predominantly internal for both polarizations. The simulated DF spectra reproduce this differential response well, confirming that the substrate selectively modifies the more externally extended ED mode while leaving the MD comparatively robust.

The CL response on bulk Si substrate differs from that on the Si$_3$N$_4$ membrane. At 30\,kV [Fig.~\ref{fig:fig4}\pnl{e}], experiment and simulation are in excellent agreement: the position-resolved map reproduces the same set of dipole and quadrupole resonances, consistent with the substrate-modified mode structure. As for the membrane-supported sphere, a feature around $\lambda\sim 400$\,nm remains visible in the experiment but is absent from the electrodynamic model, suggesting again an additional electron-induced emission channel beyond the particle eigenmodes. At 5\,kV [Fig.~\ref{fig:fig4}\pnl{d}], the experiment loses the modal spatial selectivity seen at 30\,kV, indicating that electron scattering broadens the excitation volume and washes out the impact-parameter dependence. In contrast to the Si$_3$N$_4$ case, however, the simulation still captures excitation of all major modes even at low voltage, although the straight-line-current model continues to predict a spatial selectivity that is not observed experimentally. This reduced voltage dependence is not straightforwardly explained by passive substrate effects alone: beam broadening inside the NP should still occur at 5\,kV, and the stronger localization of the ED field near the particle--substrate interface would, if anything, make the ED more sensitive to loss of beam collimation. The persistence of the ED signal at low voltage therefore suggests that on bulk Si the substrate provides an additional coupling pathway that sustains excitation of the resonant modes even when beam broadening suppresses local selectivity possibly through interface-mediated fields generated by electrons entering the Si substrate, which can couple back to the particle modes from below. Because the ED field extends much more strongly into the substrate than the MD field, such a substrate-mediated pathway can preferentially sustain ED excitation. In this sense, the Si substrate acts not only as a passive perturbation of the mode structure~\cite{vandeGroep2013,Staude2017}, but also as an active participant in the CL excitation process~\cite{Ebel:2026}.

\subsection{Reflecting metallic mirror: \uppercase{A}\lowercase{u}}

We finally consider the Si sphere on an optically thick Au substrate, shown in Fig.~\ref{fig:fig5}\pnl{a}. In this particle-on-mirror geometry, the substrate no longer acts as a weak perturbation, but hybridizes strongly with the sphere modes through image-dipole coupling. As a result, the modal response becomes strongly polarization-dependent and qualitatively different from that on Si$_3$N$_4$ and bulk Si. In particular, the mirror lifts the equivalence between in-plane and out-of-plane dipolar components and gives rise to substrate-induced hybrid modes with distinct radiative properties~\cite{Miroshnichenko2015,Sinev2016}. For the out-of-plane electric dipole, coupling to the image dipole produces a strongly gap-confined ED$_\perp$ mode with broadband response and enhanced electric field localization at the particle--metal interface~\cite{Kuo2016,Sugimoto2018}. For the out-of-plane magnetic dipole, the opposite image-dipole symmetry suppresses radiative losses and can narrow the resonance substantially, yielding the high-$Q$ MD$_\perp$ mode reported for Si particles on metallic films~\cite{Sinev2016,Sugimoto2018}.

This strong symmetry breaking is immediately evident in the DF spectra in Fig.~\ref{fig:fig5}\pnl{c}. Under s-polarized excitation, which predominantly addresses the in-plane electric dipole ED$_\parallel$, as well as MD$_\perp$, the spectrum shows a narrow peak near $\lambda\sim 620$\,nm, assigned to MD$_\perp$~\cite{Sinev2016}. This mode is narrowed by its interaction with the mirror image, whose antiparallel configuration suppresses radiative losses and produces the characteristic high-$Q$ response~\cite{Sinev2016,Sugimoto2018}. Under p-polarized excitation, which drives the in-plane magnetic dipole MD$_\parallel$ and also ED$_\perp$, the narrow peak is replaced by a broad feature extending from $\lambda\sim 550$\,nm to beyond 800\,nm. This broad response corresponds to the ED$_\perp$ gap mode, in which the out-of-plane electric dipole couples strongly to its parallel image dipole and to dissipative substrate channels, including Ohmic losses and surface plasmon polaritons~\cite{Kuo2016,Sugimoto2018}. The electrodynamic simulation reproduces both the s-polarized narrowing and the p-polarized broadening. In the unpolarized DF spectrum, the broad ED$_\perp$ contribution overwhelms the narrower MD feature, so that the electric mode appears not as a distinct peak but rather as a broad continuum.

The calculated field distributions in Fig.~\ref{fig:fig5}\pnl{b} support this interpretation. For s-polarization, the MD field retains the characteristic internal circulating character, while the corresponding image-induced field topology reflects the mirror symmetry of the metal substrate. For p-polarization, the ED field is concentrated strongly in the particle--metal gap, with enhancement factors exceeding $10^2$, consistent with the hybrid dielectric--metal nanoantenna picture developed for Si particles on mirrors~\cite{Sugimoto2018}. These maps make clear that the Au substrate does not merely shift the existing Mie resonances, but creates new hybrid modes with strong gap localization and modified radiation properties.

The CL response in Fig.~\ref{fig:fig5}\pnl{d,e} also reveals differences of mode response to the presence of mirror substrate. At 30\,kV [Fig.~\ref{fig:fig5}\pnl{e}], position-resolved spectra show a pronounced peak near $\lambda\sim 580$\,nm, attributable to ED$_\parallel$, together with features near $\lambda \sim 620$\,nm and 650\,nm associated with MD-related modes and the onset of the broad ED$_\perp$ response. Thus, CL resolves modes that are hidden in DF. In particular, the mode ED$_\parallel$ near $\lambda\sim 580$\,nm is not clearly observed in DF because its excitation requires an electric-field component normal to the substrate, which is not efficiently provided in the present DF geometry~\cite{Kuo2016}. CL, by contrast, naturally supplies such a field component through the electron trajectory and can therefore excite this mode efficiently. As a result, a resonance that is nearly inaccessible in DF becomes directly visible in CL.

At 5\,kV in Fig.~\ref{fig:fig5}\pnl{d}, the experiment again shows reduced spatial selectivity compared with 30\,kV, as on the other substrates. At the same time, the sharp mode attributed to MD$_\perp$ becomes even more prominent at low voltage. The simulations reproduce the overall spectral structure and also show a pronounced peak for excitation near the sphere center that is not observed experimentally. This discrepancy may indicate an additional channel in the calculation that does not contribute to the collected signal, possibly Si interband-related emission or strongly confined gap-plasmon-like modes at the particle--Au interface that are not efficiently outcoupled into the measured CL channel. Still, the main qualitative conclusion remains unchanged: on Au, CL reveals hybrid electric and magnetic modes that are only partially accessible in DF, and in particular makes the gap-enhanced electric resonance directly visible.

\section{Conclusions}

The combined DF and CL measurements on the same Si nanospheres across three substrates show that the observable modal spectrum is not solely an intrinsic property of the particle, but results from the interplay of substrate-induced mode modification and probe-specific excitation pathways. On the weakly perturbing Si$_3$N$_4$ membrane, the far-field response remains close to the isolated-sphere limit, while CL reveals that at low beam energy mode selectivity is strongly degraded by electron-beam broadening---an effect not captured by straight-line beam models~\cite{GarciadeAbajo1999,Stamatopoulou2024,GarciadeAbajo2021}. On bulk Si, the substrate selectively enhances the ED response in DF and CL, indicating that the substrate provides an additional excitation pathway beyond passive reshaping of the mode structure. On Au, mirror-induced hybridization creates qualitatively new particle-on-mirror modes: DF emphasizes the polarization-dependent redistribution into narrow magnetic and broad electric gap modes, whereas CL directly reveals resonances that are weak or hidden in DF. Overall, our results show that neither DF nor CL alone provides a complete picture of substrate-supported dielectric resonators. A reliable interpretation of Mie resonances on substrates therefore requires the combined use of both techniques together with electrodynamic modeling that accounts not only for the substrate, but also for the actual excitation process.

\section*{Methods}

\textbf{Si spheres synthesis}. Silicon nanospheres were synthesized following the procedure in Ref.~\cite{Sugimoto2020}, being prepared via a high-temperature reduction process, yielding colloidally dispersible particles with narrow size distributions. The resulting nanospheres were subsequently dispersed in a suitable solvent and purified to remove residual byproducts and aggregates. This synthesis route produces smooth, nearly spherical particles with high crystallinity, enabling low optical losses in the visible spectral range.

\textbf{DF spectroscopy}. Dark-field scattering spectra were acquired on a Zeiss Axio Observer microscope equipped with a halogen lamp for broadband illumination. Both excitation and collection were performed through a Zeiss EC Epiplan-Apochromat ($100\times$, $\mathrm{NA}=0.95$) HD DIC objective. The scattered light was dispersed by an Andor Kymera 193i spectrograph (100\,lines/mm grating) and recorded with a thermoelectrically cooled Andor iDus CCD camera (DU934P-BEX2-DD). For each particle, three measurements were taken under identical conditions for each illumination polarization: a sample spectrum, a background spectrum from an adjacent empty area on the same substrate, and a reference spectrum from a barium sulfate (BaSO$_4$) standard that captures the spectral profile of the lamp convolved with the wavelength-dependent throughput of the optical path. The background-subtracted sample spectrum was divided by the BaSO$_4$ reference to correct for the system response and then normalized to its maximum, yielding a dimensionless scattering spectrum suitable for cross-comparison across substrates and with simulations. The s- and p-polarized spectra were acquired by inserting a linear polarizer in the illumination path and masking the dark-field condenser so that only a 10$^\circ$ angular sector of the annular illumination remained unblocked [see Figs.~\ref{fig:fig3}\pnl{a}, \ref{fig:fig4}\pnl{a}, and \ref{fig:fig5}\pnl{a}]. 

\textbf{CL spectroscopy}. Measurements were performed in a TESCAN MIRA3 scanning electron microscope equipped with a Schottky field-emission gun and a SPARC spectral detection system (Delmic). Acceleration voltages of 5\,kV (corresponding to $v\simeq 0.14c$) and 30\,kV (corresponding to $v\simeq 0.33c$) were employed with a beam current of ca. 1\,nA, measured using a Faraday cup, which produce qualitatively different electron trajectories inside a Si sphere according to Monte Carlo simulations~\cite{Drouin2007}.
The CL signal was collected by an aluminum parabolic mirror with an aperture for the electron beam, providing an effective numerical aperture of $\mathrm{NA}=0.97$. 
The collected emission was directed to a spectrograph (Andor Kymera 193i) equipped with a 150\,lines/mm grating (500\,nm blaze) and detected with a thermoelectrically cooled Andor iDus CCD camera (DU920P-BEX2-DD-2E3).
All CL spectra were background-corrected using dark counts acquired with the electron beam blanked and subsequently corrected for the system response function of the instrument~\cite{Ebel:2025}.

\textbf{Theory}. 
The light scattering and cathodoluminescence from free-standing Si nanoparticles were calculated using Mie theory~\cite{Bohren1998,Stamatopoulou:2025}. For substrate-supported nanoparticles, we numerically calculated light scattering using the boundary-element method (BEM)~\cite{nanobem:2022,nanobem:2024} and whereas cathodoluminescence spectra was calculated using the finite-element method (FEM) within a commercially available code (COMSOL Multiphysics)~\cite{COMSOL}. 
We simulate the experimental DF spectra acquired through an $\mathrm{NA} = 0.95$ objective by calculating the scattered signal integrated over a solid angle defined by $\theta \in [0,\theta_\mathrm{i}]$ and $\phi \in [0,2\pi]$, considering a plane-wave excitation impinging the sample at $\theta_\mathrm{i} = \arcsin{\mathrm{NA}} \approx 71.8\si{\degree}$. Cathodoluminescence is obtained modeling the electron beam as a straight line-current source along a straight electron trajectory. This approximation is valid at high beam energies but becomes less accurate at low voltages in high-index materials~\cite{GarciadeAbajo2021, Stamatopoulou2024}. Distributions of electron trajectories are simulated with the Monte Carlo method using a freely available code (CASINO)~\cite{Drouin2007}.

\section*{Funding}
The Center for Polariton-driven Light-- Matter Interactions (POLIMA) is funded by the Danish National Research Foundation (Project No.~DNRF165).

P.~A.~D.~G. is supported by the Villum Young Investigator program of the VILLUM Foundation (Project No.~72118).

C.~T. acknowledges support from the Independent Research Fund Denmark (Grant No. 5281-00155B).

H.~S. and M.~F. are supported by JST FOREST Program (Grant No.~JPMJFR213L), JSPS KAKENHI (Grant No.~24K01287, 24KF0158, 25K01608, and 25K22206), and Kobe University Strategic International Collaborative Research Grant (Type B).

\section*{Acknowledgments}

The authors gratefully acknowledge Eduardo Dias, Torgom~Yezekyan, Wenhua~Zhao, and Kurt~Busch for valuable discussions and early computational efforts that helped guide the initial stages of the experimental work.

\section*{Author contributions} 

Y.~L. and S.~M. conceived the project. The Si spheres were synthesized by H.~S. and M.~F., while Y.~L. performed the experiments with assistance from S.~M. Simulations and theoretical analysis were performed by H.~H., P.~A.~D.~G., and C.~T. The initial draft was written by Y.~L. with all authors contributing to the final manuscript. S.~M. and N.~A.~M. supervised the project.
\section*{Data availability} 

Data supporting the findings of this study are available from the corresponding authors upon reasonable request.

\bibliography{bibliography}
\bibliographystyle{unsrt}

\end{document}